\documentclass[prb,twocolumn,amsmath,amssymb,superscriptaddress]{revtex4}
\usepackage{graphicx}
\usepackage{dcolumn}
\usepackage{bm}
\usepackage{epsfig}

\begin{document}
\title{Magnetic order and disorder in a quasi-two-dimensional quantum
  Heisenberg antiferromagnet with randomized exchange}

\author{F. Xiao}
\affiliation{Laboratory for Neutron Scattering, Paul Scherrer Institut, CH-5232 Villigen PSI, Switzerland}
\affiliation{Department of Chemistry and Biochemistry, University of Bern, CH-3012 Bern, Switzerland}

\author{W.J.A. Blackmore}
\affiliation{Department of Physics, University of Warwick, Coventry, CV4 7AL, UK}

\author{B.M. Huddart}
\affiliation{Durham University, Department of Physics, South Road,
  Durham, DH1 3LE, UK}

\author{M. Gomil\v{s}ek}
\affiliation{Jo\u{z}ef Stefan Institute,
Jamova c. 39, SI-1000 Ljubljana, Slovenia}
\affiliation{Durham University, Department of Physics, South Road,
  Durham, DH1 3LE, UK}

\author{T.J. Hicken}
\affiliation{Durham University, Department of Physics, South Road,
  Durham, DH1 3LE, UK}

\author{C. Baines}
\affiliation{Laboratory for Muon Spin Spectroscopy, Paul Scherrer
  Institut, CH-5232 Villigen PSI, Switzerland}

\author{P.J. Baker}
\affiliation{ISIS Pulsed Neutron and Muon Facility, STFC Rutherford
  Appleton Laboratory, Harwell Oxford, Didcot, OX11 OQX, UK}

\author{F.L. Pratt}
\affiliation{ISIS Pulsed Neutron and Muon Facility, STFC Rutherford
  Appleton Laboratory, Harwell Oxford, Didcot, OX11 OQX, UK}

\author{S.J. Blundell}
\affiliation{Oxford University Department of Physics, Clarendon Laboratory, Parks
Road, Oxford, OX1~3PU, United Kingdom}

\author{H. Lu}
\affiliation{NHMFL, Los Alamos National Laboratory, Los Alamos, NM 87545, USA}

\author{J. Singleton}
\affiliation{NHMFL, Los Alamos National Laboratory, Los Alamos, NM 87545, USA}

\author{D. Gawryluk}

\affiliation{Laboratory for Multiscale Materials Experiments, Paul Scherrer Institut, CH-5232 Villigen PSI, Switzerland}

\author{M.M. Turnbull}
\affiliation{Carlson School of Chemistry and Biochemistry and Department of
Physics, Clark University, Worcester, Massachusetts 01610, USA }

\author{K.W. Kr\"{a}mer}
\affiliation{Department of Chemistry and Biochemistry, University of Bern, CH-3012 Bern, Switzerland}

\author{P.A. Goddard}\email{p.goddard@warwick.ac.uk}
\affiliation{Department of Physics, University of Warwick, Coventry, CV4 7AL, UK}

\author{T. Lancaster}\email{tom.lancaster@durham.ac.uk}
\affiliation{Durham University, Department of Physics, South Road,
  Durham, DH1 3LE, UK}

\begin{abstract}
We present an investigation of the effect of randomizing exchange coupling
strengths in the $S=1/2$ square lattice quasi-two-dimensional quantum Heisenberg
antiferromagnet (QHAF)  (QuinH)$_2$Cu(Cl$_{x}$Br$_{1-x}$)$_{4}\cdot$2H$_2$O 
(QuinH $=$ Quinolinium, C$_9$H$_8$N$^+$), with $0\leq x \leq
1$. Pulsed-field
magnetization measurements allow us to estimate an effective
in-plane exchange strength $J$ in a regime where exchange fosters short-range order,
while the temperature $T_{\mathrm{N}}$ at which  long-range  
order (LRO) occurs is found using muon-spin
relaxation, allowing us to construct a phase diagram for the series. We
evaluate the effectiveness of disorder in suppressing $T_{\mathrm{N}}$ and
the ordered moment
size and find an extended
disordered phase in the region $0.4 \lesssim x \lesssim 0.8$ where no
magnetic order
occurs.
The observed critical substitution levels are accounted for by an
energetics-based competition between different local magnetic
orders. Furthermore, we demonstrate experimentally that the
ground 
state disorder
is driven by quantum effects of the
exchange randomness, which is a feature that has been predicted
theoretically and has implications for other
disordered quasi-two-dimensional QHAFs.
\end{abstract}
\maketitle

\section{Introduction}

Understanding the effect of disorder on magnetic ground states at a
microscopic level is an important prerequisite for future applications
of quantum-spin systems, and is the topic of a broad range of research  
(see e.g.\ \onlinecite{back1,back2,back3,back4,back5}). 
Ground states of unfrustrated magnets with classical moments are predicted to
be robust with respect to low levels of disorder, while such disorder
is thought to have a far stronger effect on quantum spin systems \cite{sandvik,two,three,seven,footnote,sandvik2}. The
two-dimensional (2D) $S = 1/2$ square lattice quantum Heisenberg
antiferromagnet (QHAF) has previously been investigated
in this context through introduction of nonmagnetic on-site impurities in
CuO \cite{one} and CuF$_{4}$ \cite{KCuFe1, KCuFe2, KCuFe3}
planes. However, less work exists on other forms of
quenched disorder such as randomized exchange bonds, where the
strength of exchange coupling is varied throughout the lattice. Numerical
treatments of this problem \cite{two} suggest that if the bond
disorder is homogeneous, the ground state is very
robust, even against strong bond disorder, with the spin stiffness and
order parameter being exponentially reduced and only vanishing in the case
of infinite randomness. However, if disorder is inhomogeneous
\cite{three,sandvik2} the occurrence of lower-dimensional quantum states, such
as dimer singlets, significantly enhances quantum
fluctuations, which reflect low-temperature time-dependence in the
states of the system (and differ from  time-independent,
temperature-driven classical fluctuations that dominate magnetism
at elevated temperatures). Disorder can also give rise
to spin frustration which strongly suppresses correlation
lengths \cite{seven,otto}.
In these
cases, long-range order can be destroyed, with a quantum-disordered
phase resulting \cite{sandvik2,kawamura, mat_sug}. 
We present here a complete experimental investigation of a 2D QHAF with
randomized exchange strengths. We indeed find evidence for 
formation of small clusters of fluctuating quantum spins acting to
destabilize magnetic order. 

\begin{figure}
\includegraphics[width=\columnwidth]{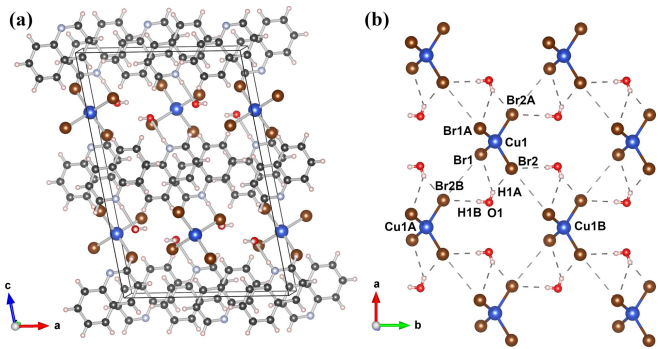}
\caption{(a) Packing diagram for (QuinH)$_{2}$CuBr$_{4}\cdot
  2$H$_{2}$O showing magnetic layers separated by quinolinium
  cations; (b) Layers of CuBr$_{4}^{2-}$ distorted tetrahedra.
\label{fig:structure}}
\end{figure}

We use coordination chemistry
to generate a tuneable family of low-dimensional materials in
which  $S = 1/2$ Cu$^{2+}$ ions are linked magnetically via a
superexchange
pathway mediated by halide bonds. Previous work shows
that by substituting halide ions in the superexchange pathway,
differing exchange strengths can be realised \cite{Schlueter,
  Liu, thede}. 
% For example,
%an investigation of one-dimensional
%Cu(py)$_{2}$(Br$_{1-x}$Cl$_{x}$)$_{2}$ (py=pyridine, C$_{5}$H$_{5}$N)
%\cite{}, showed that randomized bonds have a strong
%effect on the exchange energy $J$, ordering temperature
%$T_{\textsc{N}}$ and sublattice magnetization $m$.
The square lattice case is addressed here through pulsed-field
magnetization and muon-spin relaxation ($\mu^{+}$SR) measurements of the
series (QuinH)$_{2}$Cu(Cl$_{x}$Br$_{1-x}$)$_{4}\cdot$2H$_{2}$O
(QuinH=Quinolinium, C$_{9}$H$_{8}$N$^{+}$) \cite{lynch,Butcher,landee_new,si}.
This combination
of techniques is well suited to
determining the magnetic ground state of low-dimensional Cu$^{2+}$
complexes
\cite{Goddard,Goddard2016,landee}. 
Our series is based on
2D antiferromagnetic (AF) layers of Cu$Z_{4}^{2-}$ distorted
tetrahedra (where the halide $Z$ = Cl or Br). Tetrahedra are
related by C-centering, resulting in a square magnetic
lattice, with each $S = 1/2$ Cu$^{2+}$ ion having four identical nearest
neighbours. Hydrogen bonding to water molecules within the layer
generates close $Z$--$Z$ contacts, providing the
AF superexchange
pathway [Fig.~\ref{fig:structure}(b)].
These 2D
AF layers are well isolated owing to
the presence of alternating layers of QuinH cations 
[Fig~\ref{fig:structure}(a)]. 
The
magnetic properties of the $x = 0$ compound
(QuinH)$_{2}$CuBr$_{4}\cdot$2H$_{2}$O suggest it represents a good
realization of the 2D QHAF model with intraplane exchange strength
$J(x = 0) = 6.17(3)$~K \cite{Butcher}. Comparing $x = 1$ ($Z$ = Cl)
and $x = 0$ ($Z$ = Br) materials, there are differences of
only 
4$\%$ and 0.4$\%$ respectively in the distance between Cu$^{2+}$  ions
along the $a$-axis and $b$-axis.
 However, the change in the interaction strength caused by the varying chemical composition of the superexchange pathways will have a much larger effect than these small differences would suggest.
%We expect that these differences will have a much smaller effect on
%the magnetism than that caused by the varying chemical composition of
%the superexchange pathways.
 Energy-dispersive X-ray spectroscopy (EDX)
measurements \cite{si} were used to determine $x$ and 
confirm that there is no macroscopic separation of Br- and Cl-rich
structures.

\section{Results}
\subsection{Magnetometry}
\begin{figure*}
\includegraphics[width=17cm]{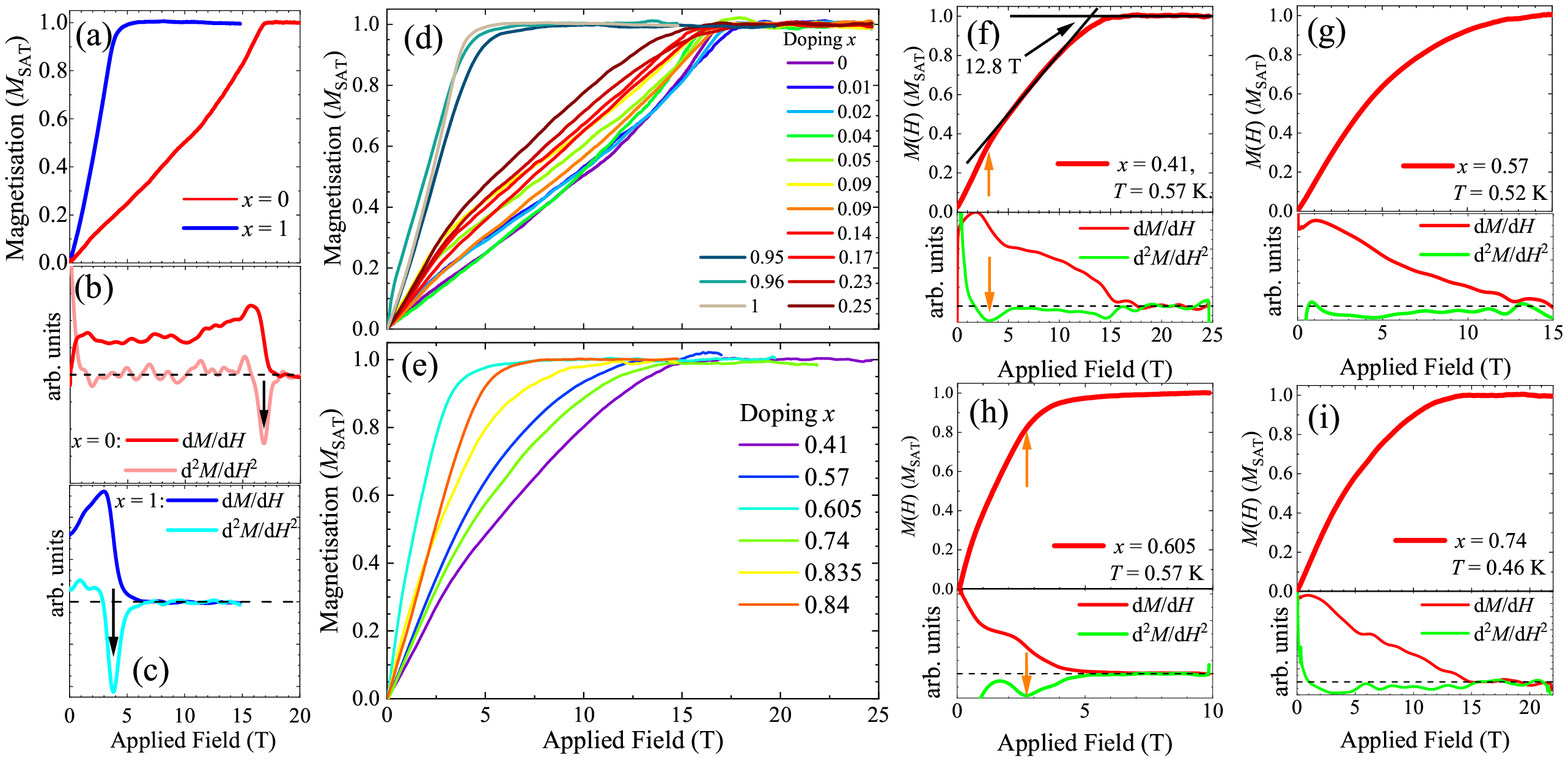}
\caption{
%  Low-temperature ($T\approx 0.6$~K) single-crystal magnetization data.
 % (a)-(c) $M(H)$, ${\rm d}M/{\rm
%  d}H$ and ${\rm d}^2M/{\rm d}H^2$ for $x = 0$ and $x = 1$. (d) Low
%and high values of $x$ show a sharp change in the slope of $M(H)$ at $H_{\rm
 % sat}$, 
%but (e) intermediate values do not. Details of the data for
%(f) $x = 0.41$ and (g) $x = 0.74$. Black arrows indicate $H_{\rm  sat}$,
%orange arrow indicates the low-field kink feature.
Low-temperature ($T \approx 0.6$ K) single-crystal magnetization data
for (QuinH)$_{2}$Cu(Cl$_{x}$Br$_{1-x}$)$_{4}$. (a)-(c) M(H), d$M/$d$H$
and d$^{2}M/$d$H^{2}$ for $x = 0$ and $x = 1$. (d) Low and high values
of $x$ show a sharp feature in $M(H)$ at $H_{\textrm{sat}}$, but (e)
intermediate values do not. Data for (f) $x = 0.41$, (g) $x = 0.57$,
(h) $x= 0.605$ and (i) $x = 0.74$ showing the smooth approach to
saturation in the intermediate values of $x$. Black arrows indicate
$H_{\textrm{sat}}$, horizontal black dashed lines correspond to zero
values of the derivatives, and orange arrows indicates the low- field
kink feature discussed in the text. Data for all samples are provided
in the Supplemental Material (\onlinecite{si}).
  \label{fig:magnetometry}}
\end{figure*}

To determine the effective intraplane exchange $J$, low-temperature
($T \approx 0.6$~K) pulsed-field magnetization 
measurements were made on materials with $0 \leq x \leq 1$
(Fig.~\ref{fig:magnetometry}) (see also the Supplemental
Material~\cite{si} where the full data set is presented along with
further details
of the analysis).
Magnetization measurements are made at
$T \ll J$ where collective behaviour of the spins is
expected. The magnetization $M$ as a function of applied field
 for the $x = 0$ and 1 materials [Fig.~\ref{fig:magnetometry}(a)] shows a convex
rise to saturation, indicative of 2D magnetic interactions
\cite{Goddard}. Where sufficient correlations (promoted by a narrow
distribution in $J$) are present (see below), saturation
of $M$ at applied field $H_{\mathrm{sat}}$ occurs via a sharp change in
the slope of $M$, giving rise to a minimum in d$^{2}M/$d$H^{2}$ that allows $H_{\mathrm{sat}}$
to be determined. For $x = 0$, this occurs at $\mu_{0}H_{\mathrm{sat}} = 16.9(4)$~T,
whereas for $x = 1$ we find  $\mu_{0}H_{\mathrm{sat}} = 3.8(3)$~T
[Fig.~\ref{fig:magnetometry}(b) and (c)].
The Hamiltonian that  describes the two end members of the family is
\begin{equation}
{\cal H} =J\sum_{\langle i,j \rangle_\parallel}{\bf S}_i\cdot {\bf S}_j +
J_\perp\sum_{\langle i,j \rangle_\perp}{\bf S}_i\cdot {\bf S}_j 
-g\mu_{\rm B}B\sum_iS_i^z,
\end{equation}
where $J$ is the strength of the exchange coupling within the magnetic
planes, $J_\perp$ is the coupling between planes, and $J \gg
J_\perp$. The first two terms on the right hand side refer to
summations over unique exchange bonds parallel and perpendicular to
the planes, respectively. For $S = 1/2$ spins, within a mean-field
treatment of this model\cite{Goddard}, saturation occurs when $g\mu_{\textsc{B}}\mu_{0}H_{\mathrm{sat}} =
zJ$ and $z = 4$ is the number of nearest neighbours in
the 2D planes. Using the published value of $g = 2.15$ for the $x = 0$
material \cite{Butcher}, this gives $J(x = 0) = 6.1(1)$~K, in good
agreement with the previous estimate. Assuming a similar $g$-factor, a
value of $J(x = 1) = 1.4(1)$~K is obtained, in good agreement
with the value derived from susceptibility
measurements\cite{landee_new} and  consistent with previous
measurements that suggest $J_{\textrm{Br}} \approx 4J_{\textrm{Cl}}$
for Cu$^{2+}$ QHAFs \cite{Schlueter}. 

For concentrations with $x \gtrsim 0$ we again measure the
characteristic 2D convex rise to saturation, but this becomes less
pronounced for $x \geq 0.05$ where the saturation field [and
therefore $J(x)$] decreases and the change in the slope of $M(H)$ becomes less sharp [Fig.~\ref{fig:magnetometry}~(d)]. As
$x$ is increased further towards $x \approx 0.4$, the approach to
saturation broadens, such that the trough in
d$^{2}M/$d$H^{2}$ is hard to discern \cite{si}. However, a sharp elbow in $M(H)$
is still observed at the saturation field, which can be identified by
extrapolation of the data above and below $H_{\mathrm{sat}}$
[Fig.~\ref{fig:magnetometry}(f)].
For $x =
0.57, 0.74$ and $0.835$ there is no clear feature in the $M(H)$ data \cite{si}
and it is
not possible to estimate an effective value for
$J$ [Fig.~\ref{fig:magnetometry}(e) and (g--i)].
In this region $M(H)$ no longer exhibits its convex
form, but instead rises smoothly with decreasing gradient up to
saturation. This behaviour is reminiscent of a disordered system,
however the data cannot be fitted to a fully paramagnetic model. This
suggests that, while interactions between spins exist, correlations
characterized by a single effective exchange energy are not present,
or drop below a certain critical length scale.
%In this region $M(H)$ rises smoothly to saturation
%but cannot be fitted to a Brillouin function, suggesting that, while
%interactions between spins exist, correlations characterized by a
%single effective exchange energy are not present, or drop below a
%certain critical length scale. 
The sharp change in the slope of $M(H)$ at
saturation  becomes  resolvable again for $x \geq
0.84$ and, as the concentration approaches $x = 1$, the traces develop
the convex shape observed at low $x$. This is consistent with the
return to 2D QHAF behavior in the $x = 1$ material.   

We can assess the coherence length $\xi$ required to give a resolvable
transition in $M(H)$ through temperature-dependent pulsed-field measurements of
the $x = 0$ compound between 0.5 and 15~K, shown in the Supplemental
Material \cite{si}. As $T$ is
raised, the saturation point becomes more rounded such that the width
of the trough in d$^{2}M/$d$H^{2}$ increases and the amplitude
decreases. For $T \gtrsim 4$~K it is no longer
possible to clearly identify $H_{\mathrm{sat}}$. The coherence length
in square lattice
planes can be estimated using $\xi/d \approx 0.498(1 –
0.44T/J)$exp$(1.131J/ T)$ where $d$ is the
magnetic lattice parameter \cite{twelve}, which holds for $H=0$ and
$T\ll J$.
Coupling this formula with the limiting value of $T$, above which
$H_{\mathrm{sat}}$ is undefined, suggests that the magnitude of
exchange can be identified only when $\xi/d
\gtrsim 2$ at $H=0$.  

In addition to the feature at saturation, the $M(H)$ data for some
samples show a kink at fields considerably lower than $H_{\rm
  sat}$ for the $x = 1$ system. The kink is resolvable
%[Fig.~\ref{fig:magnetometry}(d)]
for several $x$ between 0.05 and 0.61, indicated by an orange arrow in
Figs.~\ref{fig:magnetometry}(f) and (h). We attribute this to
the presence of isolated clusters of spins (e.g.\ dimers, trimers, 
square plaquettes, etc.) coupled by Cl--Cl halide exchange bonds, which are weaker than Br--Br bonds and thus easier to saturate with an applied field. (The effect
of these localized units is discussed below.)

\subsection{Muon-spin relaxation}

\begin{figure*} 
\includegraphics[width=14cm]{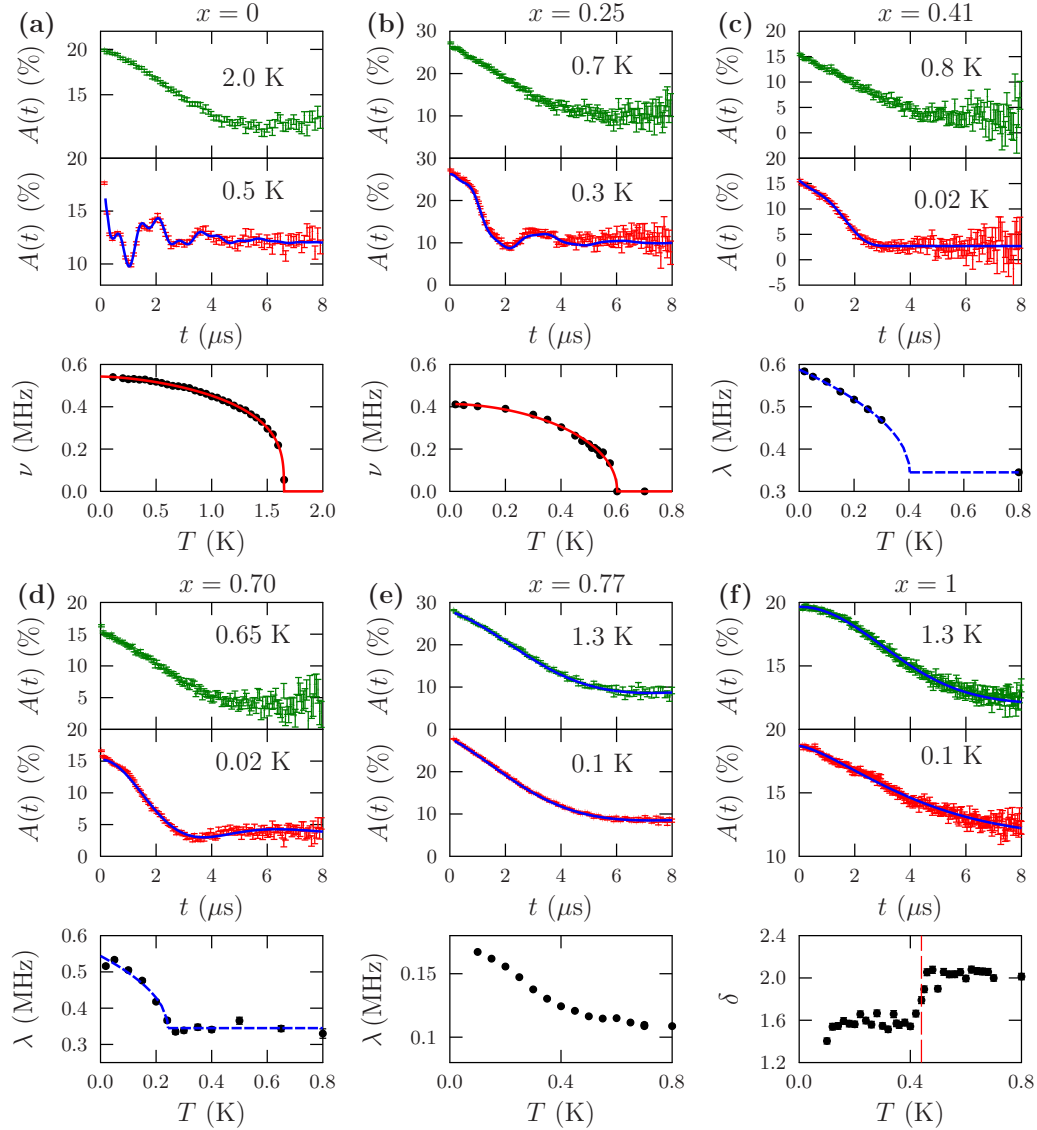}
\caption{Results of ZF $\mu^{+}$SR measurements. (a-f) {\it Top}: Example high-temperature  spectra; {\it
    Middle}: example low-temperature spectra; {\it Bottom}: Example
  oscillation frequency $\nu$, relaxation rate $\lambda$ or stretching
  parameter $\delta$ from the function ${\rm e}^{-(\lambda t)^{\delta}}$
  (see main text). 
\label{fig:muonfig}}
\end{figure*}

Although the ideal 2D QHAF should only show  long-range magnetic order
(LRO) at $T=0$, 
in any  realization of the model in a three-dimensional material the presence of interplane exchange 
$J_{\perp}$ can lead to
a transition with
$T_{\mathrm{N}}>0$. To determine $T_{\mathrm{N}}$,  
zero-field (ZF)
$\mu^{+}$SR measurements were made \cite{steve,yaouanc}.
Oscillations in the asymmetry are observed in some members of the
series at low $T$ (Fig.~\ref{fig:muonfig}), providing
unambiguous evidence  of
LRO.
For materials with $x\leq 0.25$ oscillations are observed at multiple ($n=2$ or $n=3$)
frequencies $\nu_{i}$ 
[Fig.~\ref{fig:muonfig}(a,b)] consistent with several magnetically
inequivalent muon sites.
The oscillatory spectra can be fitted to a function of the form
\begin{equation}
A(t)=\sum_{i=1}^{n}A_ie^{-\lambda_it}\cos(2\pi\nu_it+\phi_i)+
A_\mathrm{bg}e^{-\lambda_\mathrm{bg}t}.\label{eqn:asymmetry-fit} 
\end{equation}
where the last term accounts for muons with their initial spin polarization
along the direction of the local magnetic field, along with those muons that
stop in the sample holder. The frequencies were held in fixed
proportion for the fits (fitting parameters are given in the
Supplemental Material \cite{si}) From the behaviour of the oscillatory frequency versus temperature,
the ordering temperature $T_{\mathrm{N}}$ for each of the compounds can be extracted using
the function $\nu_i(T)=\nu_i(0)\left[1-\left(T/T_{\mathrm{N}}\right)^\alpha\right]^\beta$,
which provides values consistent with discontinuous
changes in amplitude that also occur at the ordering transition.
We find 
$T_{\mathrm{N}}(x=0)=1.65(1)$~K, and
transition temperatures that decrease smoothly with increasing
$x$, such that
 $T_{\mathrm{N}}(x)$ extrapolates to zero at $x\approx
0.35$. 
The frequencies $\nu_{i}(T\rightarrow 0)$ are proportional to the
moment size on the Cu$^{2+}$ ions and hence to the sublattice
magnetization $m$.  We measure relatively small
frequencies compared to typical 3D systems,
reflecting a reduced ordered moment (expected to be
0.33$\mu_{\mathrm{B}}$ for $T\rightarrow 0$ in spin wave
theory \cite{manou}).
 These frequencies decrease with increasing $x$
with $m$ dropping by around 24\% from $x=0$ to $0.25$.  

The behavior is qualitatively different for samples with $0.41 \leq x
\leq 0.77$ [Fig.~\ref{fig:muonfig}(c--e)] where no oscillations are
resolved down to 0.02~K.
Instead, spectra resemble a distorted Kubo-Toyabe (KT)
function\cite{yaouanc} at low $T$,
corresponding to
disordered quasistatic moments in the materials,
with the distortion of the spectra likely reflecting short-range order
along with some limited dynamic fluctuations. As $T$ is
increased, the spectra change such that they resemble dynamic, exponential
functions above $T\gtrsim 0.5$~K.
These data can be parametrized using a stretched-exponential envelope function $e^{-(\lambda t)^{\delta}}$ that accounts for
the early time behaviour of the spectra.
The transition between the static and dynamic regimes appears
abrupt in the $x=0.70$ sample, taking place at a freezing
temperature to a glassy configuration around  $T_{\mathrm{f}}=0.27$~K, with a similarly rapid variation in
relaxation rate seen in the $x=0.41$ material at low temperature, suggesting $T_{\mathrm{f}}\approx
0.41$~K. No such sharp freezing is seen in the  $x=0.77$ sample, where the
relaxation rate $\lambda$ drops fairly smoothly with increasing $T$
(with a change in slope around $T=0.4$~K, likely related to the freezing
seen for other concentrations).

\begin{figure}
\includegraphics[width=\columnwidth]{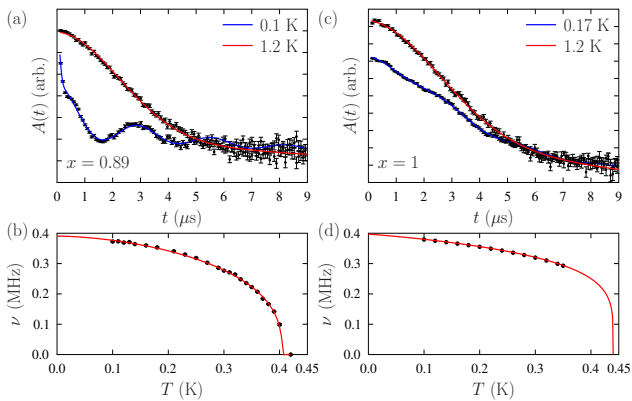}
\caption{
  (a) Muon-spin relaxation data for the
$x=0.89$ material below (0.1~K) and above (1.2~K) the ordering temperature
$T_{\mathrm{N}}$, with a fit shown to a Bessel function
relaxation (blue curve, $T<T_{\mathrm{N}}$) and relaxed Kubo-Toyabe function
(red curve $T>T_{\mathrm{N}}$). The extracted frequency is plotted in
(b), where the line is a guide to the eye.
(c) Data for the $x=1$ material below (0.17~K)
  $T_{\mathrm{N}}$ and above (1.2~K), 
  with a fit to an oscillatory
model for the low-temperature data (blue curve) and relaxed Kubo-Toyabe function
(red curve $T>T_{\mathrm{N}}$). The extracted frequency is plotted in (d), where the line is a guide to the eye.
\label{fig:muon2}}
\end{figure}

The observed behavior is qualitatively different again in the $x=0.89$
and $x=1$ materials [Fig.~\ref{fig:muon2}]
 where an abrupt transition  to LRO takes place with similar
 $T_{\mathrm{N}}$.
Data for the $x=0.89$ material [Fig.~\ref{fig:muon2}(a)] can be fitted to a Bessel function,
typical of incommensurate magnetic order \cite{yaouanc}. The
presence of incommensurate order might also be consistent with measured
data
for $0.1\leq x < 0.41$ where non-zero phase offsets are observed in
the oscillatory components, although the presence of multiple characteristic
frequencies complicates the modelling of this feature.
The Bessel
function results from sampling a distribution of local magnetic fields
that varies sinusoidally with position in the material, as expected from an
incommensurate  spin-density wave. However, depending on the muon sites in a system,
there are other field distributions that can 
lead to relaxation that resembles the Bessel functional form, with its
characteristic negative phase shift and damped cosinusoidal temperature
dependence. As a result, it is not possible to unambiguously infer the existence
of an incommensurate magnetic structure in this composition. In any
case, the characteristic frequency decreases smoothly [Fig.~\ref{fig:muon2}(b)] allowing
$T_{\mathrm{N}}=0.41(1)$~K to be extracted using the same approach as for the
materials with $x\leq 0.25$.

Data for the $x=1$ composition show oscillations below the ordering
temperature, but at relatively low 
amplitude compared to other concentrations,
as shown in Fig.~\ref{fig:muon2}(c). The frequency
of these oscillations varies smoothly with
temperature [Fig.~\ref{fig:muon2}(d)], but cannot be reliably fitted to
an oscillatory function close to the transition, where the relaxation
rate increases. Such low-amplitude oscillations have been observed
previously  in
similar materials with related structures \cite{steele,lancaster}. In
this case, the $x=1$ crystallites are notably different in surface
colour and form to the
other concentrations, and the relatively large ratio of relaxing to
oscillatory signal, could
 reflect the behaviour of muons near the surfaces of these
 crystallites. However, the transition is via a discontinuous
 change in the spectra (also seen in the other compositions, where it
 coincides with the disappearance of the oscillations) and this
 feature is used to assign 
$T_{\mathrm{N}}(x=1)=0.44(1)$~K.

% This is seen via $\mu^{+}$SR oscillations at a
% single frequency  in the $x=0.89$ material, and via
% the $\beta$ parameter
%in the $x=1$ material, where oscillations occur only with low
%amplitude \cite{si}.

For the $x=0$ material, we have $T_{\mathrm{N}}/J=0.27(2)$, which  combined with
predictions from Quantum Monte Carlo (QMC) simulations
\cite{2dformula}, suggests
$|J_{\perp}/J|\approx3.2\times 10^{-3}$, indicating well-isolated magnetic layers.
At  $x=1$  we observe magnetic order with
$T_{\mathrm{N}}/J=0.31(2)$ and thus $|J_{\perp}/J| \approx 7.5\times 10^{-3}$. Comparing, we have $J_{\perp}(x=1) = 0.014(5)$~K and
$J_{\perp}(x=0) = 0.011(8)$~K, which is the same within
uncertainties, 
demonstrating that
the degree of isolation of the 2D layers is largely unaffected by
substitution of Br for Cl ions. This implies that $J_{\perp}(0<x<1)$ is likely
close to these values, and that the observed magnetic effects of bond randomness are
attributable solely to  disorder in the 2D layers.

\section{Discussion}

\begin{figure} 
\includegraphics[width=\columnwidth]{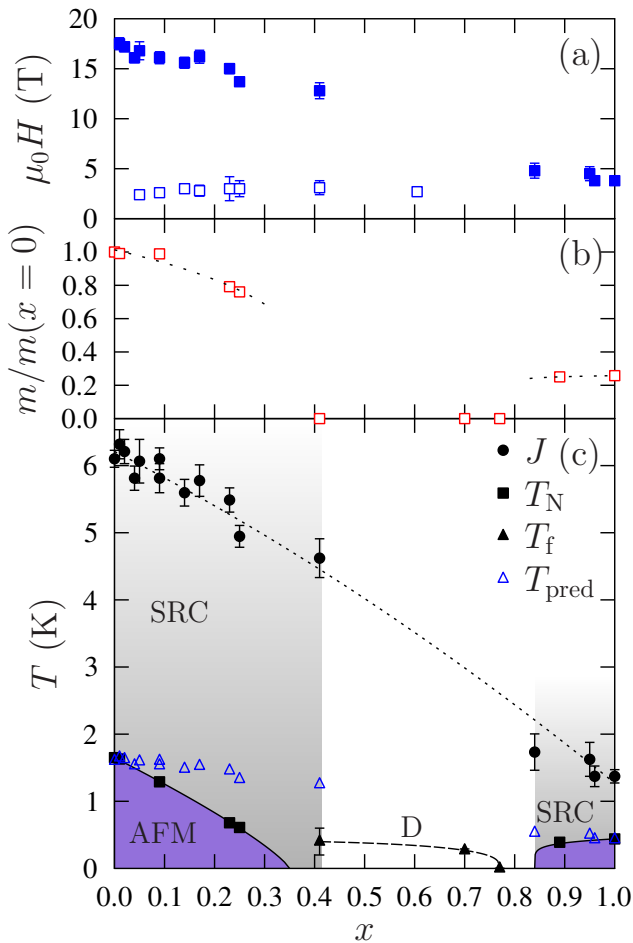}
\caption{(a) Fields at which $\mu_{0}H_{\mathrm{sat}}$  (filled
  symbols)  and low-field kink (open symbols)  are observed as a
  function of $x$.
  (b) Evolution of  estimated ordered moment.
(c) Notional phase diagram showing antiferromagnetically ordered (AFM), short-range
correlated (SRC) and disordered (D) regions.
%The dashed line is a 
 % crossover from the D region suggested by data measured for
 % $x=0.61$.
Open triangles show the predicted ordering temperatures from QMC 
  assuming $J_{\perp}=0.011$~K (and no disorder effects aside from a renormalized effective $J$). Dotted line described in main text.
\label{fig:phasediagram}}
\end{figure}

A notional phase diagram for the system  is shown in
Fig.~\ref{fig:phasediagram}.
The parameter $x$ represents the
fraction of Cl in a square 2D unit cell with intermediate values
corresponding to more exchange-bond disorder. 
Since
halide bonds are formed
from two $Z$ ions, the presence of Cl can create a
Cl--Cl exchange bond [expected to be around 4 times weaker than Br--Br
bond exchange based on the size of $J(x)$] or a mixed Cl--Br
bond.
The effective exchange strength $J$ extracted from  $M(H)$
data provides the energy scale below which we 
expect short-range AF correlations in 2D planes 
to dominate the magnetic behaviour for
$T_{\mathrm{N}}\ll T \ll J$.
The phase diagram is not symmetrical
about $x=0.5$ because $x$ does not merely
lead to random substitution but also decreases the
effective value of $J$ across the series.

We expect the effective exchange $J$ through halide-halide contacts to
reflect the size and shape of the orbitals.
Structurally, the exchange strength $J$ via the two-halide pathway 
depends on the identity of the halide ion for two
reasons.  The first is the shape of the orbitals which leads
to better overlap between bromides than between chlorides. The second
is that
the inter-halide distance is shorter for
Cl.
By
substituting Cl for Br at low levels of doping, the cell constants
will still be similar to the $x=0$ compound, so that not only is the Cl ion
smaller leading to poor overlap, but the distance between the Cl and Br
may be greater than observed in the pure Br material, leading to a still smaller
value of $J$.  At low concentrations of Br the lattice is
similar to the $x=1$ material and the opposite trend might be expected, with Br ions
in small spaces causing distortion, and
therefore with shorter than expected halide-halide distances, leading
to a larger value of $J$. 

The extracted values of $J(x)$ [Fig.~\ref{fig:phasediagram}(c)] show a gradual
decrease up to $x= 0.41$.
This is also the region where LRO
is observed, with $T_{\mathrm{N}}$ showing a similar gradient to
$J(x)$.
Combining the measured $J(x)$ with our estimated $J_{\perp}$ we
can use the QMC results \cite{2dformula} to predict values of
$T_{\mathrm{N}}$ assuming disorder leads only to a renormalized effective $J$ (open triangles in
Fig.~\ref{fig:phasediagram}).
The measured $T_{\mathrm{N}}$ are seen to depart significantly from these
predictions, showing that disorder does have a strong effect in suppressing
$T_{\mathrm{N}}$ beyond simply the gradual reduction in effective $J$.
The ordered moment is seen to decrease
as shown in Fig.~\ref{fig:phasediagram}(b).
The behaviour in this part of the phase diagram is reminiscent of that for substitutional
disorder in La$_{2}$Cu$_{1-z}$(Zn,Mg)$_{z}$O$_{4}$ \cite{one}. A
fairly linear decrease was observed in $T_{\mathrm{N}}$ and the
ordered moment, along with the disappearance of LRO
around $x=0.41$. There is also  resemblance to the 1D molecular case
 in Ref.~\onlinecite{thede} where $J$ values change approximately
linearly across the phase diagram,
while $T_{\mathrm{N}}$ and ordered moments drop rapidly on
the Br-rich side of the phase diagram.
In our case, the energy scales close to $x=1$ are all lower owing to a smaller
$J$ mediated by the Cl ions.
In the region $0.84 \leq x \leq 1$ there is a sufficient correlation
length to identify $J$ from the $M(H)$ data and LRO is restored above
$x = 0.89$. However, $T_{\mathrm{N}}$ close to $x=1$ does not show the
rapid decrease seen on the other side of the phase diagram when moving
away from the pristine composition, likely because enhanced disorder
is also accompanied by an increase in effective $J$.

For $0.41 \leq x \leq 0.84$ the magnetic behaviour is more
complicated.
No LRO can be
identified from the $\mu^{+}$SR data across the entire region. 
The lack of a sharp feature in $M(H)$ at saturation implies that collective behaviour
characterized by a single effective exchange $J$ is no longer
straightforwardly applicable and that there is therefore a highly magnetically disordered
region. Here we see
 evidence from $\mu^{+}$SR for slow fluctuations of
 spins for $T \gtrsim 0.5$~K with these becoming more static
at the lowest measured temperature, although still not long-range
ordered down to 0.02~K. The lack of muon oscillations in the static regime
points\cite{yaouanc} to a coherence length $\xi/d\ll 10$.
% while the lack of $M_{\mathrm{sat}}$ suggests $\xi/d \lesssim 2$.
Non-zero $M$ at small applied field
implies that this
disordered phase is not characterized by an energy gap. 
For samples with $ 0.41\leq x \leq 0.7$ there is also evidence for 
freezing of spins at low $T$. 
This would appear to suggest freezing of glassy behaviour in this region, as might
be expected for a system forming clusters of strongly interacting spins
surrounded by disordered moments \cite{binder}, and seems to be
distinct from the spin-liquid-like state predicted for random
interactions \cite{sandvik2}.

\begin{figure}
\includegraphics[width=\columnwidth]{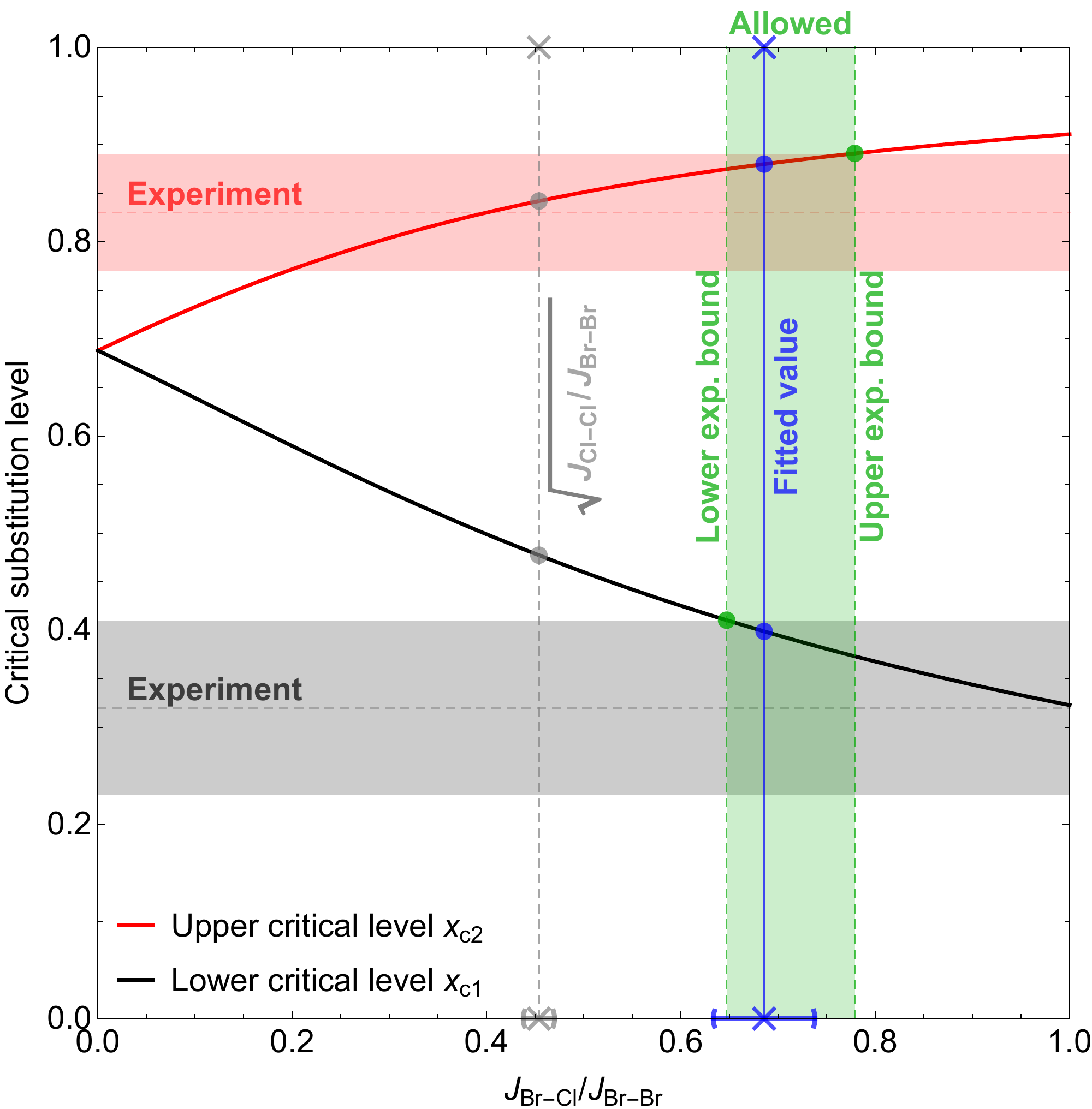}
\caption{Lower ($x_{c1}$, black solid line) and upper ($x_{c2}$, red
  solid line) critical substitution levels calculated from the
  bond-energetics criterion for magnetic-order collapse detailed in
  the main text as a function of the ratio
  $J_\text{Br--Cl}/J_\text{Br--Br}$. (The equations determining these
  substitution levels
  are identical when $J_\text{Br--Cl}$ vanishes, resulting in a single critical
  value.) The measured ratio $J_\text{Cl--Cl}/J_\text{Br--Br} = 0.206(15)$ is
  assumed. Experimentally-observed critical substitution levels
  $x_{c1}$ and $x_{c2}$ are indicated by horizontal dashed lines with
  one-sigma shaded regions around them. The green shaded region
  highlights the range of ratios $J_\text{Br--Cl}/J_\text{Br--Br}$
  compatible with $x_{\mathrm{c}1}$ and $x_{\mathrm{c}2}$ within one
  sigma. The simple guess $J_\text{Br--Cl}/J_\text{Br--Br} \approx
  \sqrt{J_\text{Cl--Cl}/J_\text{Br--Br}}$ is shown as a grey
  vertical dashed line and lies outside the experimentally-allowed
  green region, while the actual value of
  $J_\text{Br--Cl}/J_\text{Br--Br}$ extracted from the best fit of the
  average exchange $J(x)$ [dashed line in Fig.~\ref{fig:phasediagram}(c)] is shown by a
  blue vertical line and is compatible with experimental
  $x_{\mathrm{c}1}$ and $x_{\mathrm{c}2}$. One-sigma
  uncertainty intervals for both of these values are shown by
  horizontal error bars on the bottom axis.} 
\label{sifig1} 
\end{figure}

We consider here three potential effects driving the form of the phase diagram: (i) percolation;
(ii) bond energetics and (iii) quantum fluctuations. 
The bond percolation threshold for a square lattice is \cite{stauffer} $p_{\mathrm{c}}=1/2$. However, for
our materials a single exchange bond comprises two possible
substitution sites.
If a single substitution per bond suffices to destabilize magnetic
order then we should equate the percolation threshold $p_{\mathrm{c}}$ to the
probability that one or more substitutions occurs on a single
exchange bond  $p_{\mathrm{c}} = 1 - (1 - x_{\mathrm{c}1})^2$, which gives a
lower critical substitution level $x_{\mathrm{c}1}  = 0.29$, while at
high $x$,
we should have $p_{\mathrm{c}} = 1 - x_{\mathrm{c}2}^2$,
which gives an upper critical substitution level $x_{\mathrm{c}2} = 1 - x_{\mathrm{c}1} = 0.71$. This could
be compatible with the data for $x<0.41$, but fails to describe the
large-$x$ behavior. Furthermore, it is unlikely that
percolation is the sole driver of the observed behaviour since we are
changing the strengths of random bonds, rather than
removing exchange pathways. More sophisticated correlated percolation models
including lattice-dependent grouping of substituted bonds also fail to
describe the measured phase diagram. The possibility of such clusters
forming, their size and the effect of correlated substitutions  are
discussed in Appendix~\ref{sect:clusters}.

An approximate criterion for the collapse of magnetic order (which
could be short range)
might be when the total exchange energy of  substituted bonds
becomes larger than that of  unsubstituted bonds.
We would expect a lower critical substitution level $x = x_{\mathrm{c}1}$ to be determined by
Br--Cl and Cl--Cl bonds acting as disorder in a Br--Br ordered
background such that $(1 - x_{\mathrm{c}1})^2 J_{\mathrm{Br}-\mathrm{Br}}  =
 2x_{\mathrm{c}1}(1-x_{\mathrm{c}1}) J_{\mathrm{Br}-\mathrm{Cl}} +
x_{\mathrm{c}1}^{2}J_{\mathrm{Cl}-\mathrm{Cl}}$. 
The upper critical substitution level $x = x_{\mathrm{c}2}$ is then determined by
Br--Cl and Br--Br bonds acting as disorder in a Cl--Cl ordered
background giving $x_{\mathrm{c}2}^2 J_{\mathrm{Cl}-\mathrm{Cl}} =
2x_{\mathrm{c}2}(1-x_{\mathrm{c}2})J_{\mathrm{Br}-\mathrm{Cl}} +
(1-x_{\mathrm{c}2})^{2}J_{\mathrm{Br}-\mathrm{Br}}$.
The unknown exchange strength in these expressions, $J_\text{Br--Cl}$, can
be determined by fitting the measured $J(x)$
with $J(x) = (1-x)^2 J_\text{Br--Br} + 2x(1-x)
J_\text{Br--Cl} + x^2 J_\text{Cl--Cl}$, which describes the
data  well [dotted line, Fig.~\ref{fig:phasediagram}(c)] and gives estimates
$J_\text{Br--Br} = 6.2(1)~\mathrm{K}$, $J_\text{Br--Cl} =
4.3(3)~\mathrm{K}$, and $J_\text{Cl--Cl} = 1.3(1)~\mathrm{K}$. These
yield the critical substitution levels $x_\mathrm{c1} = 0.40(2)$ and
$x_\mathrm{c2} = 0.88(2)$, 
both of which agree well with the observed location of
the collapse of
magnetic order.

In fact, values of $x_{\mathrm{c}}$ compatible with
experiment result
from only a limited range of choices for the ratio
$J_\text{Br--Cl} /J_\text{Br--Br}$.
%The bond-energetics criterion for the collapse of magnetic order
%detailed in the main text allows us
We can express the expected lower and
upper critical substitution levels $x_{c1}$ and $x_{c2}$,
respectively, as a function of the two exchange-strength ratios
$J_\text{Br--Cl}/J_\text{Br--Br}$ and
$J_\text{Cl--Cl}/J_\text{Br--Br}$.
Fixing the known pristine-system
exchange-strength ratio $J_\text{Cl--Cl}/J_\text{Br--Br} = 0.206(15)$,
the observed critical substitution levels $x_{c1} = 0.32(9)$ and
$x_{c2} = 0.83(6)$ put stringent limits on the range of
experimentally allowed ratios $J_\text{Br--Cl}/J_\text{Br--Br} =
0.65\text{--}0.78$ (Fig.~\ref{sifig1}). This is incompatible with the most
simple assumption that each substitution of a Br with a Cl ion
weakens the exchange bond (which consists of two Br/Cl sites) by the
same factor, which would yield $J_\text{Br--Cl}/J_\text{Br--Br}
\approx \sqrt{J_\text{Cl--Cl}/J_\text{Br--Br}} = 0.45(2)$. On the
other hand, the  enhanced ratio $J_\text{Br--Cl}/J_\text{Br--Br}
= 0.69(5)$ extracted from the best fit of the average exchange model
to the measured $J(x)$
[Fig.~\ref{fig:phasediagram}(c)] is fully compatible with the
observed critical substitution levels via the bond-energetics
criterion (Fig.~\ref{sifig1}). This validates both the
bond-energetics criterion for the collapse of magnetic order in the 2D
square-lattice QHAF as well as the average-exchange $J(x)$ model
described above.

Finally, theory predicts that the disorder-driven introduction of
antiferromagnetically-coupled dimers, chains or other clusters
acts to enhance quantum fluctuations,  destroying long-range magnetic order
\cite{three,sandvik2}. This scenario is consistent with our observations: the presence
of the low-field kink in our magnetometry data points to high
densities of microscopic clusters of Cu moments coupled by Cl bonds,
while our EDX measurements showed no evidence for phase separation,
suggesting
inhomogeneities are limited to a local
level. Calculations indeed show (Appendix~\ref{sect:clusters}) that a random distribution of disordered bonds leads to a
large concentration of dimers and trimers
around $x=0.2$, where we see $T_{\mathrm{N}}$ being
strongly suppressed towards disorder.

\section{Conclusion}

In summary, 
%we have performed a complete experimental investigation of
%the effect of bond substitution in the 
 the addition of small
amounts of disorder to the pristine 2D QHAF causes regions of
the sample to remain correlated with a single effective $J$,
which decreases as $x$ increases. Simultaneously there is a
preponderance for  formation of minority clusters (e.g.\ dimers and
trimers) that enhance quantum fluctuations and act to suppress $T_{\rm
  N}$ more than is predicted from the change in $J$ alone. For
$0.41 \leq x \leq 0.84$, while spins continue to interact,
the correlated regions are no longer apparent, LRO is
completely absent and low-temperature spin freezing is evident.
Critical substitution levels can be explained by an energetics-based competition between different local magnetic orders.
%Critical
%substitution levels can be explained using arguments based on bond
%energetics.
Our result that magnetic order can be destroyed by quantum effects of
exchange randomness could have implications for other disordered Q2D
AFM systems such as the parent state of the cuprate superconductors,
or frustrated square lattices, which are believed to evolve into a
spin-liquid state on the introduction of quenched disorder.

\acknowledgments
Part of this work was carried out at 
S$\mu$S, Paul Scherrer Institut, Switzerland and STFC-ISIS Facility,
Rutherford Appleton Laboratory, UK. We are grateful to EPSRC (UK) for
financial support. 
This project is supported by the
European Research Council (ERC) under the European Union’s Horizon
2020 research and innovation program (Grant Agreement
No.~681260).
FX thanks C.P.\ Landee for inspiring discussion and B.\ Frey for assistance with  EDX measurements in Bern.  
WJAB thanks the EPSRC for additional funding. Work at the
National High Magnetic Field Laboratory is supported by NSF
Cooperative Agreements No. DMR-1157490 and No. DMR-1644779, the State
of Florida, the US DOE, and the DOE Basic Energy Science Field Work
Project Science in 100 T. The financial support by the Swiss National
Science Foundation under grant no.\ 200020\_172659 is gratefully
acknowledged. MG thanks the Slovenian Research Agency for additional funding under project No.\ Z1-1852.
Data presented here will be made available via
Durham Collections.

\appendix

\section{Calculating the effect of halide substitution on spin clusters}
\label{sect:clusters}

In the main text we described the influence of percolation as a
possible driver
of the phase diagram of this system. The complication of
exchange bonds comprising two possible substitution sites makes this
problem more complex than that of a bond being formed from a
single substitution sites. We discuss the details of the effects of
halide substitutions on spin clusters in this Appendix.

\subsection{Formation of spin clusters}
\begin{figure}
\includegraphics[width=\columnwidth]{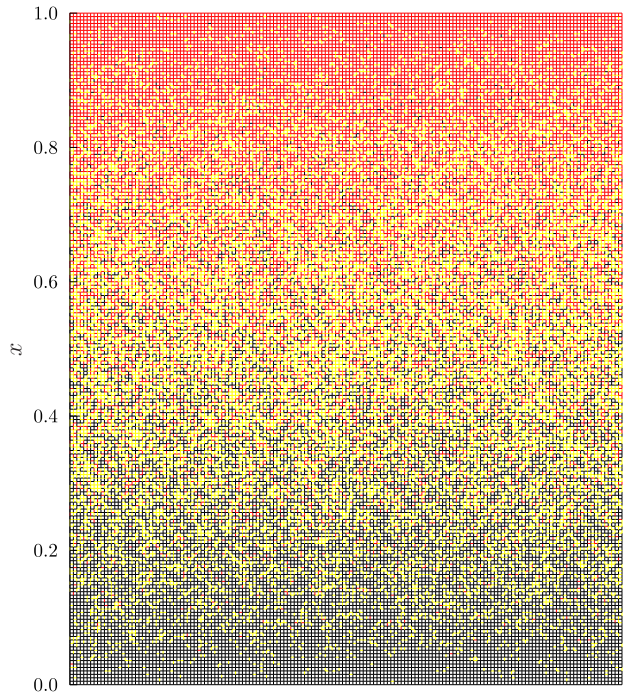}
\caption{A schematic diagram showing a simulation of the
  two-dimensional lattice in (QuinH)$_2$Cu(Cl$_x$Br$_{1-x}$)$_4\cdot
  2$H$_2$O for different values of $x$.  The coloring of the bonds is
  black for Br--Br, yellow for Br--Cl, red for Cl--Cl.
}
\label{fig:lattice}
\end{figure}

Increasing $x$ in (QuinH)$_2$Cu(Cl$_x$Br$_{1-x}$)$_4\cdot 2$H$_2$O has
the result of replacing Br linkages with Cl linkages in superexchange pathways.  Since bonds are
formed with two halide ions, the connections between magnetic
Cu$^{2+}$ ions change from being all Br--Br links at $x=0$ (colored
black in Fig.~\ref{fig:lattice}) to all Cl--Cl links at $x=1$ (colored
red).  However, at intermediate $x$ there are many Br--Cl links
(colored yellow).  This is shown schematically in
Fig.~\ref{fig:lattice} in which the bonds are chosen randomly
according to the value of $x$ indicated on the vertical axis.
This figure illustrates that intermediate values of $x$ give a range
of mixtures of different linkages which, as explained in the main
text, have different exchange
strengths.

\begin{figure}
\centering
\includegraphics[width=\columnwidth]{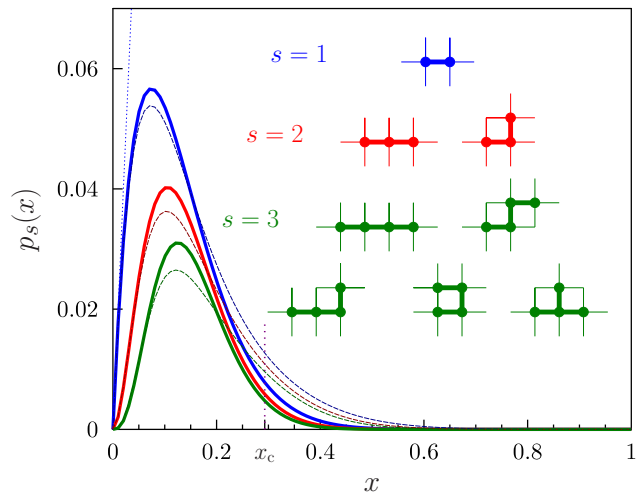}
\caption{The probability of isolated clusters of Cl-containing bonds
  [dimers (blue),
    trimers (red) or tetramers (green), corresponding to $s=1$, $2$
    and $3$ bond clusters] surrounded by only Br--Br bonds. The dotted line 
  tangent to the $s=1$ curve is
  described in the text.  The percolation threshold for impurity bonds
  with at least one substitution $x_{\rm c}$ is 
    indicated by the short vertical dotted line.
  The dashed curves show the effect of including $\pm 30$\%
  additional inhomogeneous
  clustering of Br- or Cl-rich regions.
}
\label{fig:quin}
\end{figure}

Another way of looking at this problem is shown in Fig.~\ref{fig:quin}
which plots the probability of finding an isolated dimer, trimer or
tetramer of spins connected by  Cl-containing bonds (either Br--Cl or
Cl--Cl) and surrounded
by only Br--Br bonds that dominate at low $x$. The probability that a randomly chosen bond contains at least one Cl is $1-(1-x)^2 = x(2-x)$. Furthermore, we denote by
 $p_s(x)$ the probability that a randomly chosen bond 
is part of an isolated cluster of Cl-containing bonds of size $s$ (so
a spin dimer has $s = 1$ because it represents two Cu$^{2+}$ spins
connected by a single Cl-containing bond, a spin trimer has $s = 2$
because it involves two Cl-containing bonds, etc.).  These
probabilities are given by $p_{1}(x) = x(2-x)(1-x)^{12}$ for spin
dimers
[the power of 12 reflecting the six double-bromide bonds that must be present at the boundary of a dimer,
a double Br bond having probability $(1-x)^2$],
$p_{2}(x) = 6x^2(2-x)^2(1-x)^{16}$ for spin trimers (both straight and bent, involving
eight double-bromide bonds at the boundary) and
$p_{3} = 3x^3(2-x)^3(1-x)^{18}(9(1-x)^2+2)$ for tetramers (where various
shapes are possible, as shown in Fig.~\ref{fig:quin}).  For very small
$x$ these values depend mostly on the probability of finding enough Cl-containing
bonds (so for dimers, this factor is $1-(1-x)^2=x(2-x)$, the dotted
blue line in Fig.~\ref{fig:quin}) but this becomes reduced at higher
$x$ due to the probability of finding pure Br--Br bonds surrounding
the cluster falling substantially below unity.
The result is that these isolated species are only reasonably probable
below the ideal percolation threshold for impurity bonds with at least
one Cl substitution, which is $x = x_{\rm c} = 1 - 1/\sqrt{2}\approx 0.29$ [a value obtained\cite{stauffer}
from the exact bond percolation threshold for a square lattice in
terms of the single bond occupation probability $p_{\mathrm{c}} = 1/2 = x_{\mathrm{c}}(2 - x_{\mathrm{c}})$
]. Beyond this value clusters start to link up, and 
  these species are practically absent
above $x\approx 0.4$.  This concentration is roughly consistent with
the lower value
of $x=x_{\mathrm{c}1}$ at which both the AFM and SRC phases collapse, although this likely reflects the fact that
$x\approx 0.4$ lies well above the percolation threshold. This is
explored in more detail below. 
% $p_{\mathrm{c}}$. 

\subsection{Size of spin clusters}

\begin{figure}
\centering
\includegraphics[width=\columnwidth]{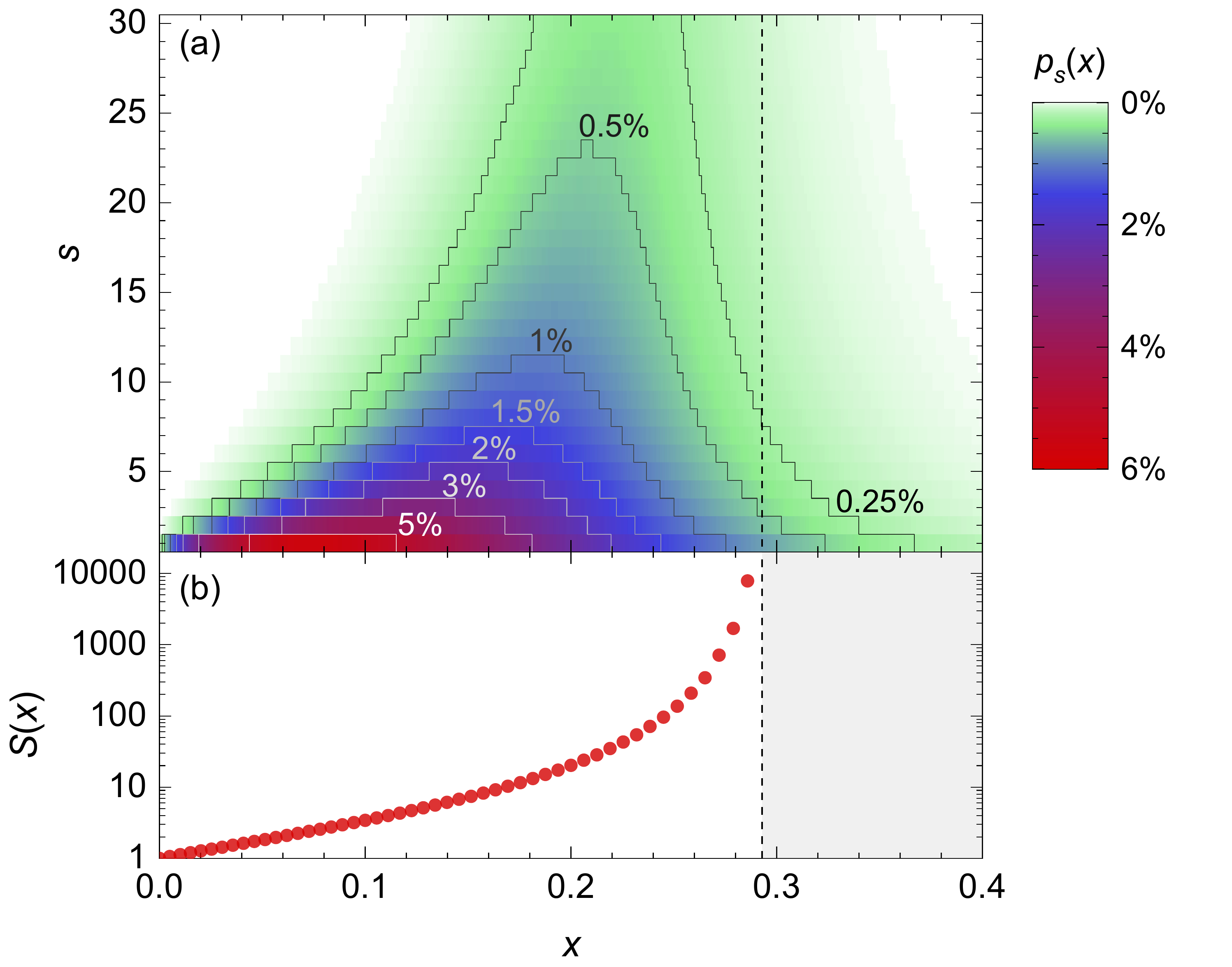}
\caption{(a) A density plot of the cluster probabilities $p_s(x)$.  (b) The mean size $S$ of
  bond clusters as a function of $x$.  The percolation threshold $x_{\mathrm{c}}$ for impurity bonds with at least one substitution is
  indicated by a vertical dashed line.
  }
\label{fig:cluster}
\end{figure}

The effect of $x$ on these isolated clusters is seen even more clearly
in Fig.~\ref{fig:cluster}(a) which shows the form of $p_s(x)$ as a
function of both $x$ and cluster size $s$.
These probabilities were computed numerically on the line graph
(i.e.\ graph of bonds) of an $L \times L$ patch of a square lattice
with $L = 4001$. An $O(L)$ space-complexity algorithm inspired by the
modified Hoshen--Kopelman
algorithm~\cite{hoshen1976percolation,tiggemann2006percolation} and
generalized to arbitrary lattices was used for this, as described in
Ref.~\onlinecite{gomilsek2019kondo}. 
Note that
$\sum_{s=1}^{\infty} p_s(x)=x(2-x)$,
the probability that one or more substitutions occurs on a single
bond, since the sum over $s$ of probabilities that a randomly chosen
bond is a member of a cluster of size $s$ necessarily accounts for all
substituted bonds (at least below the percolation threshold).
We note that the probabilities of Br-rich bond clusters in a Cl--Cl
bond background (which dominate for $x$ near 1) are the same as the ones
presented on Figs.~\ref{fig:quin} and \ref{fig:cluster} under the duality $x \rightarrow 1-x$
that exchanges the roles of Br and Cl.

Dividing $p_s(x)$ by $x(2-x)$ gives the conditional probability that a
randomly-chosen Cl-substituted
bond is part of a cluster of size $s$ and allows us to estimate the
mean cluster size $S(x)$.
This quantity is therefore defined as\cite{stauffer}  $S(x)=\sum_{s=1}^\infty s
p_s(x)/[x(2-x)]$ and
answers the question:
given a randomly chosen bond with one or more substitutions,
what is the mean size of the
cluster that this bond is a part of?  This is plotted in
Fig.~\ref{fig:cluster}(b), showing that the mean cluster size
increases rapidly as $x$ increases, and diverges as $x\to x_{\rm c}$.
Given the sharpness of this  percolation transition we
conclude that uncorrelated percolation on its own is likely not the
sole driver behind the collapse of AFM and SRC order in the material
studied in this paper. Namely, the
lower critical concentration $x_{c1} = 0.40(2)$ is not close enough to
the  sharp percolation threshold of Cl-rich bond clusters, which
occurs at a substantially lower $x_{\mathrm{c}} = 1 -
1/\sqrt{2} \approx 0.29$.
The discrepancy is even more acute for the upper critical
concentration $x_{\mathrm{c}2} = 0.88(2)$,
which is substantially higher than the predicted
percolation threshold
of Br-rich bond clusters
$1 - x_{\mathrm{c}} = 1/\sqrt{2} \approx 0.71$ under the Cl--Br
duality. Uncorrelated percolation thus cannot explain the observed
critical concentrations for the collapse of AFM and SRC order in these
systems.

\subsection{Effect of correlated substitutions}

Going beyond uncorrelated percolation, we first consider the effect of
locally-correlated substitution on individual bonds due to structural
consequences of changing the size of the halide ion (the ionic radius
of Br$^-$ is ${\approx}~8\%$ larger than that of Cl$^-$). Denoting the
probabilities that a randomly chosen bond is a Br--Br bond, a Cl--Cl
bond or a mixed Br--Cl bond, by $p_{\text{Br--Br}}$,
$p_{\text{Cl--Cl}}$ and $p_{\text{Br--Cl}}$, respectively,
uncorrelated substitutions of Br by Cl on bonds would correspond to
the probabilities: $p_{\text{Br--Br}}^0 = (1-x)^2$,
$p_{\text{Br--Cl}}^0 = 2 x (1-x)$ and $p_{\text{Cl--Cl}}^0 =
x^2$ (where the superscript $0$ labels the probability for an
uncorrelated substitution). Structural changes might skew these probabilities, but
they must obey $p_{\text{Br--Br}} +
p_{\text{Br--Cl}} + p_{\text{Cl--Cl}} = 1$, and reproduce the
observed Cl concentration by obeying $p_{\text{Br--Cl}}/2 +
p_{\text{Cl--Cl}} = x$. Denoting the $x$-dependent probability
difference ${\Delta p}(x) = p_{\text{Br--Cl}}(x) -
p_{\text{Br--Cl}}^{0}(x)$ for finding a mixed Br--Cl bond, so that ${\Delta p}
= 0$ would correspond to locally-uncorrelated substitutions, we get:
$p_{\text{Br--Br}} = (1-x)^2 - {\Delta p}/2$, $p_{\text{Br--Cl}} =
2 x (1-x) + {\Delta p}$ and $p_{\text{Cl--Cl}} = x^2 - {\Delta
  p}/2$. In this model the lower percolation threshold of Cl-rich bond
clusters would correspond to the condition: $1 -
p_{\text{Br--Br}}(x_{\mathrm{c}1}) = p_{\mathrm{c}} = 1/2$, while the upper percolation
threshold for Br-rich bond clusters would correspond to the condition:
$1 - p_{\text{Cl--Cl}}(x_{\mathrm{c}2}) = p_{\mathrm{c}} = 1/2$. Solving for the
mixed-bond probability shift ${\Delta p}$ assuming locally-correlated
bond substitutions we get ${\Delta p}(x_{\mathrm{c}1}) = -0.28(5)$ and ${\Delta
  p}(x_{\mathrm{c}2}) = +0.55(7)$, which correspond to a decrease of mixed-bond
probabilities by $-58(7)\%$ at $x = x_{\mathrm{c}1}$ and an increase of
mixed-bond probabilities by $+260(80)\%$ at $x= x_{\mathrm{c}2}$. The huge shifts in
probabilities that this model would  require
are implausible,
allowing us to reject this percolation model with purely
bond-local substitutional correlations.

We also tested the effect of including additional clustering effects
between different neighbouring bonds (i.e. a scenario of inter-bond correlated percolation) due to
the structural consequences of changing the
size of the halide ion.  This could mean that slightly Br-rich
regions and slightly Cl-rich regions could spontaneously form in a
crystal prepared with a particular nominal $x$, though we stress that
we have no experimental evidence that this effect occurs in our
samples.  To model this, we considered a sample with an equal mixture
of regions with $x(1+\epsilon)$ and with $x(1-\epsilon)$ and
illustrate the effect in Fig.~\ref{fig:quin} for $\epsilon=0.3$.
This would be an extremely high level of clustering, but the
simulations show that this does not alter the general conclusions
stated above.  The only effect observed is a small shift of the
probability of isolated dimers, trimers and tetramers to larger values of $x$
(due, of course, to regions of the sample in which $x$ is smaller than
the nominal value).  We conclude that our picture of isolated clusters
of Cl-rich bonds growing as $x$ increases, starting to coalesce and
essentially disappearing completely above around $x\approx 0.4$
is fairly robust to clustering effects and cannot explain the
conflicting experimental values of $x_{\mathrm{c}1} > x_{\mathrm{c}}$ and $x_{\mathrm{c}2} \gg 1 -
x_{\mathrm{c}}$.  

For a quantitative understanding of correlation effects we consider
the exact correlated-percolation model of Ref.~\onlinecite{gomilsek2019kondo}, where a shift of
probability that a bond is substituted by Cl if a neighbouring bond is
also substituted by Cl by some constant factor $\gamma > 0$, where
$\gamma = 1$ corresponds to uncorrelated percolation, would correspond
to an effective rescaling\cite{gomilsek2019kondo} $p \rightarrow \mathrm{min}(\gamma p_0, 1)$
of the probability that a bond contains at least one Cl, where
$p_0 = x(2-x)$ is the uncorrelated probability. Since the lower
experimental critical concentration $x_{\mathrm{c}1}$ is larger than the
uncorrelated percolation expectation of  $x_{\mathrm{c}} = 1 - 1/\sqrt{2} \approx
0.29$, we would get $\gamma = p_{\mathrm{c}}/p_0(x_{\mathrm{c}1}) = 0.78(3) < 1$. This is
quite a large deviation from uncorrelated percolation (it corresponds
to a Pearson correlation coefficient of $\phi = (\gamma - 1)/(p^{-1}_{\mathrm{c}} -
1) = -0.22(3)$ at the percolation threshold), and being less than
unity means that nearby Cl-substituted bonds are less likely than
expected for uncorrelated Cl substitutions. The effect would be that
Cl would actually be dispersed more evenly throughout the sample than
by pure uncorrelated chance. By extension, in the dual view of rare Br
bond substitutions in a Cl--Cl bond background (valid for $x \approx 1$)
one should also get less clustering of Br-rich bonds (as clustering of
Br-rich bonds would also push Cl-rich bonds closer together, rather
than further apart as required by $\gamma < 1$), meaning that the dual
upper critical concentration $x_{\mathrm{c}2}$ should get pushed to lower
values (further away from $x = 1$) than the uncorrelated expectation
of $1 - x_{\mathrm{c}} = 1/\sqrt{2} \approx 0.71$, in clear contradiction with
experiment where $x_{\mathrm{c}2} = 0.88(2) \gg 0.71$. 

We therefore conclude that a dual pair of pure percolation transitions
of Cl- and Br-rich bond clusters cannot explain the experimentally
observed critical concentrations $x_{\mathrm{c}1}$ and $x_{\mathrm{c}2}$ neither via
uncorrelated Cl substitutions, via bond-local correlation of
Cl substitutions, nor via inter-bond substitutional correlation effect
\cite{gomilsek2019kondo}.
In contrast, the experimentally observed critical concentrations are
reproduced relatively straightforwardly using a simple position-blind model of
substituted-bond energetics (see previous section). We therefore
conclude that a lattice-dependent bias towards grouping of substituted
bonds must not be particularly significant in the system that we have
studied, and is therefore not the primary driver behind the ultimate
collapse of AFM and SRC order, which most likely originates from an
energetics-based competition between different local magnetic
orders. On the other hand, the simple unbiased model confirms the
random formation of minority clusters (e.g.\ dimers and trimers) at low
substitution values, as suggested by the magnetisation
measurements. We reassert that these clusters will promote quantum
fluctuations, and in all likelihood account for the observed
suppression of $T_{\mathrm{N}}$ beyond that which would be expected from the
reduction in the effective exchange strength alone.


\begin{thebibliography}{xx}
\bibitem{footnote}
The distinction between classical and quantum in this context rests on
whether the properties of the system can be described 
using magnetic moments free to point in any direction, or if the
quantization of spin components must be considered.
\end{thebibliography}
\end{document}